\documentclass[aps,prb,twocolumn]{revtex4}
\usepackage{amsmath}
\usepackage{amssymb}
\usepackage{graphicx}

\begin{document}

\title{Effect of a thin optical Kerr medium on a Laguerre-Gaussian beam}

\author{Weiya Zhang}
\email{weiya_zhang@wsu.edu}

\author{Mark G. Kuzyk }
\email{kuz@wsu.edu}

\affiliation{Department of Physics and Astronomy, Washington State University, Pullman, WA 99164-2814}


\begin{abstract}
Using a generalized Gaussian beam decomposition method we determine the propagation of a Laguerre-Gaussian (LG) beam that has passed through a thin nonlinear optical Kerr medium. The orbital angular momentum per photon of the beam is found to be conserved while the component beams change. As an illustration of applications, we propose and demonstrate a z-scan experiment using an $LG_0^1$ beam and a dye-doped polymer film.
\end{abstract}

\maketitle

\noindent Laguerre-Gaussian (LG) modes were discovered soon after the invention of the laser\cite{Kogelnik:1966}. A LG beam with angular mode number $l$ and transverse radial mode number $p$ can be written as:
\begin{align}
LG_p^{l}\left(r,\phi,z\right) & =\left(\frac{{\omega _0 }}{{\omega\left(z\right)}}\right) \left( \frac{\sqrt 2r}{{\omega \left(z \right)}}\right)^{\left|l\right|} L_p^{\left|l\right|} \left(\frac{{2r^2 }}{{\omega ^2\left(z \right)}}\right) \label{Eq:LG_beam} \nonumber \\
&\quad \times \exp \left(\frac{{ - r^2 }}{{\omega ^2\left(z \right) }}\right) \exp \left(- i\frac{{kr^2 z }}{{2\left(z^2  + z_r^2 \right)}} \right) \nonumber \\
&\quad \times  \exp \left(i\left(2p +\left| l\right| + 1\right)\tan ^{ - 1} \left(\frac{{ z}}{z_r}\right)\right)\nonumber \\
&\quad \times \exp \left( - i{l}\phi \right)\exp \left(- ikz\right),
\end{align}
where $\omega _0$ is the beam waist radius, $z_r=k \omega _0^2/2$ is the Rayleigh length, $\omega (z) = \omega _0 (1 + z^2 /z_0^2 )^{\frac{1}{2}} $ is the beam radius at z, and $L_p^l$ is an associated Laguerre polynomial. For a given LG beam mode, $z_r$ (or $\omega _0$) alone is sufficient to characterize the relative amplitude and phase of the electric field of the beam. When multiple beams are involved, it is often necessary to specify the waist locations of each beam. For convenience we use the notation $C \cdot LG_p^{l}(r,\phi,z-z_w; z_r)$ to describe a LG beam unambiguously, where $z_w$ is the waist location on the z axis and C is a complex constant that gives the amplitude and the initial phase.

The fundamental Gaussian beam $LG_0^0$ has been the most commonly studied LG mode in both theory and experiment. However recently higher order LG beams, especially those with higher angular mode number $l$, are attracting more attention. When $l\geq1$, the LG beam possesses well defined orbital angular momentum of $l\hbar$ per photon\cite{Allen:1992}, zero on-axis intensity due to the phase singularity, and helical phase fronts.  These properties make high-order LG beams useful in many applications such as optical manipulation of microscopic particles\cite{Paterson:2001}, quantum information processing\cite{{Mair:2001},{Barreiro:2003}}, orbital angular momentum sorting\cite{Wei:2003} and imaging\cite{Torner:2005}, nonlinear wave mixing\cite{Courtial:1997}, optical vortices\cite{Rozas:1997} and vortex solitons\cite{Kivshar:1998}, scattering\cite{Schwartz:2005}, and interference\cite{Sztul:2006}.  In the present work, we develop a theory for the propagation of an LG beam after it passes through a thin nonlinear optical Kerr medium.
The results of our calculation can be used for  z-scan measurements using $LG_0^1$ beam and optical limiting geometries.

The index of refraction of an optical Kerr medium depends on the intensity $I$ of the incident beam and the nonlinear refractive index $n_2$,
\begin{align}
n=n_0+n_2I,  \label{Eq:n2}
\end{align}
where $n_0$ is the linear refractive index. Consider a nonlinear Kerr-type sample that is placed at the position $z=z_s$ in a LG beam $E(r,\phi,z)=E_0 \cdot LG_{p_0}^{l_0}(r,\phi,z;z_r)$. Assume the sample is ``thin'' such that the change of the diameter of the beam due to either linear diffraction or nonlinear refraction can be neglected within the sample. Such two  requirements are satisfied if  $d<<z_r$ and $d<<z_r/\Delta \Phi$, respectively, where d is the sample thickness, and $\Delta \Phi$ is the nonlinear phase distortion\cite{Sheik:1990}.  The electric field $E'$ at the exit surface of the sample can be expressed as\cite{Sheik:1989,Sheik:1990}
\begin{align}
E'\left(r,\phi,z_s\right)= E\left(r,\phi,z_s\right)e^{-\frac{\alpha d}{2}}e^{-i\Delta\Phi\left(r,\phi,z_s\right)}, \label{Eq:eexit}
\end{align}
where $\alpha$ is the linear absorption coefficient, and
\begin{align}
\Delta \Phi\left(r,\phi,z\right) &= \frac{\Delta\Phi _0}{1+z^2/z_r^2} \left( \frac{2r^2}{{\omega^2 \left(z \right)}}\right)^{\left|l_0\right|}   \nonumber \\
&\quad \times \left(L_{p_0}^{\left|l_0\right|} \left(\frac{{2r^2 }}{{\omega ^2\left(z \right)}}\right)\right)^2 \exp \left(\frac{{ -2 r^2 }}{{\omega ^2\left(z \right) }}\right), \label{Eq:nonlinearphase}
\end{align}
where
\begin{align} \label{Eq:phi0}
\Delta \Phi_0 = \frac{ \pi}{\lambda } c \epsilon _0 n_0 n_2 |E_0|^2\frac{1-e^{-\alpha d}}{\alpha}
\end{align}
is a constant proportional to the maximum nonlinear phase change $\Delta \Phi_{max}(z_s)$ in the sample at the focus $(z_s=0)$ . When the incident beam is a $LG_0^0$ beam, $\Delta \Phi_0 = \Delta \Phi_{max}(z_s=0)$.

We generalize the so-called ``Gaussian decomposition'' method \cite{Weaire:1979} to analyze the propagation of a beam that traverses a nonlinear sample. The exponential in Eq. (\ref{Eq:eexit}) can be expanded in a Taylor series as
\begin{align}
e^{-i\Delta\Phi\left(r,\phi,z_s\right)}=&\sum^{\infty}_{m=0} \frac{ \left( -i\Delta\Phi\left(r,\phi,z_s\right) \right) ^m}{m!}. \label{Eq:phaseexpansion}
\end{align}
The complex electric field of the incident beam after it passes through the sample can be written as a summation of the electric fields of a series of LG beams of different modes as
\begin{align}
E'\left(r,\phi,z\right)=\sum^{\infty}_{m=0} \sum_{p=0}^{p_m}C_{p,m}LG_p^{|l_0|}\left(r,\phi,z-z_{wm};z_{rm}\right), \label{Eq:Eafter}
\end{align}
where $z_{wm}$ and $z_{rm}$ are the waist location and  the Rayleigh length, respectively, of the corresponding beam mode and $C_{p,m}$ is the amplitude and phase of the component beam. These parameters are determined by letting $z=z_s$ in Eq. (\ref{Eq:Eafter}) and making use of Eqs. (\ref{Eq:eexit}) and (\ref{Eq:phaseexpansion}).
\begin{align}
z_{\omega m}=z_r\frac{4m\left(m+1\right)Z}{Z^2+\left(2m+1\right)^2}, \nonumber
\end{align}
\begin{align}
z_{rm}=z_r \frac{\left(2m+1\right)\left(Z^2+1\right)}{Z^2+\left(2m+1\right)^2}, \nonumber
\end{align}
and $C_{p,m} = D_{p,m} \cdot F_{p,m}$ where Z is defined as $Z=z_s/z_r$, and
\begin{align}
F_{p,m} &= E_0 e^{-\alpha d/2}\frac{\left(-i \Delta\Phi _0\right)^m}{m!\left(2m+1\right)}
\sqrt{\frac{\left(2m+1\right)^2+Z^2}{\left(1+Z^2\right)^{2m+1}}} \nonumber \\
&\quad \times \exp \left(-i k z_r\frac{4m\left(m+1\right)Z}{Z^2+\left(2m+1\right)^2}\right) \nonumber \\
&\quad \times \exp\left(i\left(2p_0+\left| l_0 \right | +1\right)\tan^{-1}\left(Z\right)\right) \nonumber \\
&\quad \times \exp\left(-i\left(2p+\left| l_0 \right | +1\right)\tan^{-1}\left(\frac{Z}{2m+1}\right)\right), \nonumber
\end{align}
 and $p_m$ and $D_{p,m}$ are determined through
\begin{align}
\sum_{p=0}^{p_m}D_{p,m} \cdot L_p^{\left| l_0 \right |}\left(x\right) = \frac{x^{m\left| l_0 \right |}\left(L_{p_0}^{\left| l_0 \right |}\left(\frac{x}{2m+1}\right)\right)^{2m+1}}{\left(2m+1\right)^{\frac{2m+1}{2}\left| l_0 \right |}}, \nonumber
\end{align}
where x is an arbitrary real variable.  Our theoretical results are embodied in Eq. (\ref{Eq:Eafter}).

It is worth noting that all the component LG beams have the same angular mode number $l_0$ as that of the incident beam,  which reflects the conservation of the photon's orbital angular momentum. Therefore the effect of the Kerr material on the incident LG beam is to generate LG beams of different radial modes. These results are important in applications that leverage mode sensitivity.

The expression in Eq. (\ref{Eq:Eafter}) is especially useful when the nonlinear phase distortion is small enough so that only a few terms in  the summation are needed to make a good approximation. To illustrate, assume that the incident beam is a $LG_0^1$ beam and the maximum nonlinear phase distortion in the sample at position Z is
\begin{align}
|\Delta \Phi _{max}\left(Z\right)|=\frac{|\Delta \Phi _0|}{e \cdot \left(1+Z^2\right)} << 1.
\end{align}
It is sufficient to keep the first two terms in the Eq. (\ref{Eq:phaseexpansion}) and neglect the higher order terms, yielding
\begin{align}
E'\left(r,\phi,z\right) &\approx F_{0,0} LG_0^1\left(r,\phi,z;z_r\right) \nonumber \\
&\quad +\frac{2}{3\sqrt{3}}F_{0,1} LG_0^1\left(r,\phi,z-z_{w1};z_{r1}\right) \nonumber \\
&\quad -\frac{1}{3\sqrt{3}}F_{1,1} LG_1^1\left(r,\phi,z-z_{w1};z_{r1}\right). \label{Eq:elg01}
\end{align}
This outgoing electric field includes the generated $LG_0^1$ and $LG_1^1$ beam.

It is interesting to calculate the on-axis normalized Z-scan transmittance \cite{Sheik:1989,Sheik:1990}
\begin{align}
T\left(Z ,\Delta \Phi_0 \right) = \frac{{\left| {E' \left(r \to 0,\phi, z \to \infty \right)} \right|^2 }}{{\left| {E' \left(r \to 0,\phi, z \to \infty \right)|_{ \Delta \Phi_0  = 0}} \right|^2 }}, \label{Eq:normalized_T}
\end{align}
which  characterizes the on axis light power transmitted though a small aperture placed in the far field. Applying Eq. (\ref{Eq:elg01}) we find
\begin{align}
T\left(Z ,\Delta \Phi_0 \right) &= 1+ \frac{8Z\left(27+10Z^2-Z^4\right)}{\left(1+Z^2\right)\left(9+Z^2\right)^3}\Delta \Phi_0 \nonumber \\
&\quad+\frac{16}{\left(9+Z^2\right)^3}\Delta \Phi_0^2.\label{Eq:Toflg01}
\end{align}
The last term can be dropped if $|\Delta \Phi_0| <<1$.

The same procedures can be applied to LG beams of other modes. As a special case, when the incident beam is $LG_0^0$, we get the same result as in Ref.~\onlinecite{Sheik:1990}, i.e.

\begin{align}\label{Eq:Toflg00}
T\left(Z ,\Delta \Phi_0 \right) &= 1+ \frac{4Z}{\left(1+Z^2\right)\left(9+Z^2\right)}\Delta \Phi_0
\end{align}
when $|\Delta \Phi_0| <<1$.

\begin{figure}[htb]
\centerline{\includegraphics[width=8cm]{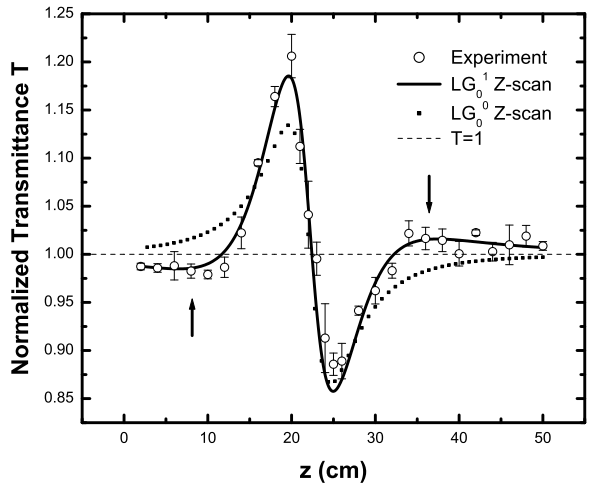}}
\caption{ \label{fig:zscan} Experimental (circles) and theoretical (solid curve) results of the Z-scan of a 1.4mm-thick 2\% DR1/PMMA sample using a $LG_0^1$ beam. Also shown are a $LG_0^0$ Z-scan trace (dotted curve) and a dashed line indicating the level of $T=1$. }
\end{figure}

Result (\ref{Eq:Toflg00}) was first reported as the theory
supporting the Z-scan technique, a highly sensitive technique to
measure the optical nonlinearities using a $LG_0^0$
beam\cite{Sheik:1990}. In a Z-scan, one measures the trace of the
normalized transmittance $T$ as a function of the nonlinear sample
position z, from which $\Delta \Phi_0$ is determined by fitting  Eq.
(\ref{Eq:Toflg00}) or more practically, through the relation $\Delta
T_{p-v}=0.406|\Delta \Phi_0|$ (derived from Eq. (\ref{Eq:Toflg00})),
where  $\Delta T_{p-v}$ is the difference between the peak (maximum)
and the valley (minimum) of the trace (e.g., the dotted curve in
Figure~\ref{fig:zscan}). $\Delta \Phi_0$ then gives $n_2$ through
Eq. (\ref{Eq:phi0}).

Similarly, Eq. (\ref{Eq:Toflg01}) shows the relationship between $T$
and $\Delta \Phi_0$, suggesting that it is possible to do a  Z-scan
experiment using the $LG_0^1$ beam to measure the nonlinearity of a
thin samples. We demonstrate this on a disperse-red-1 doped
poly(methyl methacrylate) (DR1/PMMA) thin sample\cite{Zhang:2002}.
The experiment also tests the validity of our theory.

\begin{figure}[htb]
\centerline{\includegraphics[width=8cm]{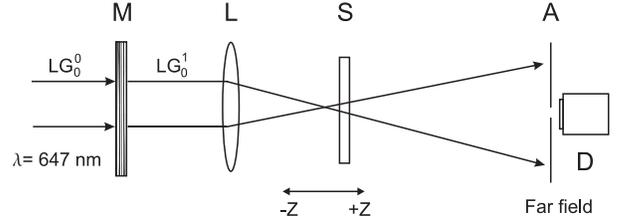}}
\caption{ \label{fig:setup} Schematic diagram of the Z-scan experiment using a $LG_0^1$ beam. M: computer generated phase mask, L: lens, S: sample,  A: aperture, and D: detector.}
\end{figure}

Figure~\ref{fig:setup} shows a schematic diagram of the setup. A
$LG_0^0$ beam of 647 nm wavelength from a krypton laser is converted
into a $LG_0^1$ beam by a computer generated binary phase mask
\cite{Heckenberg:2002}. The $LG_0^1$ beam is then focused by a lens
to the desired Rayleigh length (2.4 cm). The power of the incident
beam is 65 $\mu$W. A 1.4 mm-thick 2\% w/w DR1/PMMA sample is placed
near the beam waist. The nonlinearity of DR1/PMMA  is due to
photo-induced trans-cis-trans isomerization of DR1 molecules
followed by reorientation in the direction perpendicular to the
polarization of the incident laser beam. It can be treated as an
optical Kerr effect when the intensity is low and the exposure time
is short enough to avoid saturation of the refractive index change.
We use a shutter (not shown) to control the total exposure time (3
seconds). A small aperture is placed on axis of the beam in the far
field. The power passed through the aperture is recorded by a
detector as a function of sample position $z$ as well as the
exposure time $t$. The normalized transmittance is obtained by
dividing the  power recorded by the detector at time $ t=3$ s  by
that at time $t=0$ s. A fresh sample spot is used for each exposure
to avoid history effects. Nonlinear absorption is determined to be
negligible with an open aperture Z-scan. The ``thin'' sample
approximation is satisfied under the above experimental conditions.

Figure~\ref{fig:zscan} shows the results.  The circles are the
experimental data. The solid curve is the best fit using Eq.
(\ref{Eq:Toflg01}) with $\Delta \Phi_0$ as the parameter.  The
dotted curve is the best fit using Eq. (\ref{Eq:Toflg00}), included
as a $LG_0^0$ Z-scan trace for comparison.  The dashed line is to
indicate the line where $T=1$.

The shape of the  Z-scan curve using the $LG_0^1$ beam differs from
the traditional one using the $LG_0^0$ beam in that the former has
an extra peak and valley (indicated in the figure by the arrows).
The extra valley brings down the tail of the major peak below $T=1$
and the extra peak brings up the tail of the major valley beyond
$T=1$. The experimental data clearly shows these features and fits 
our theory well. On the other hand, the tails in the
$LG_0^0$ Z-scan trace never cross the $T=1$ line and do not fit the
data. Thus our theory provides the only good theory, supporting its
validity. The appearance of the additional structure of our theory
is a consequence of the $LG_1^1$ component beam in Eq.
(\ref{Eq:elg01}). Although the amplitude of these features is much
smaller than the major ones, it has significance in other
applications such as optical limiting, which will be discussed in a
future publication.

Just as the traditional $LG_0^0$ Z-scan, the $LG_0^1$ Z-scan
provides sensitive measurements of optical nonlinearities. The value
of $\Delta \Phi_0$ is fitted to be $-1.1 $ (or $\Delta
\Phi_{max}(Z=0)=-0.4 $) using the data in Figure~\ref{fig:zscan}.
With this we calculate $n_2 \approx -1.3\times10^{-4}$ cm$^2$/W,
which is consistent with the value measured using the traditional
$LG_0^0$ Z-scan. Besides using the fitting method, we can also
estimate $\Delta \Phi_{max}(Z=0)$ using the difference between the
major peak and valley, $\Delta T_{p-v}=0.789|\Delta
\Phi_{max}(Z=0)|$. An important difference between the two Z-scan
measurements is that in the $LG_0^1$ Z-scan experiment the
detector is placed at the beam center where the intensity is the
weakest due to the screw phase dislocation while in the $LG_0^0$
Z-scan experiment the center intensity is the strongest. As a
result the former shows a much bigger deviation from the normal
value if any phase or intensity distortion that destroys the
symmetry of the beam profile is present. This suggests that the
$LG_0^1$ Z-scan experiment is more sensitive to changes of the
$n_2$ of the sample.

In conclusion, we have calculated the propagation of a Laguerre-Gaussian beam after it passes through a thin nonlinear optical Kerr medium.  Applications, such as the Z-scan experiment using a $LG_0^1$ beam described here, has advantages over traditional measurements using a $LG_0^0$ beam.  These results can be applied to many other thin film applications such as optical limiting and nonlinear beam interactions.

We acknowledge NSF (ECS-0354736), the Summer Doctoral Fellows Program provided by Washington State University, and Wright Patterson Air Force Base for generously supporting this work.

\end{document}